# Stratified wake of an accelerating hydrofoil

**Hadar Ben-Gida**[*]
*Faculty of Aerospace Engineering, Technion - Israel Institute of Technology, Haifa, 32000, Israel*

**Alex Liberzon**
*School of Mechanical Engineering, Tel Aviv University, Tel Aviv, 69978, Israel*

**Roi Gurka**
*School of Coastal and Marine Systems Science, Coastal Carolina University, Conway, SC, 29528, USA*

**Abstract**

Wakes of towed and self-propelled bodies in stratified fluids are significantly different from non-stratified wakes. Long time effects of stratification on the development of the wakes of bluff bodies moving at constant speed are well known. In this experimental study we demonstrate how buoyancy affects the initial growth of vortices developing in the wake of a hydrofoil accelerating from rest. Particle image velocimetry measurements were applied to characterize the wake evolution behind a NACA 0015 hydrofoil accelerating in water and for low Reynolds number and relatively strong and stably stratified fluid ($Re$=5,000, $Fr \sim O(1)$). The analysis of velocity and vorticity fields, following vortex identification and an estimate of the circulation, reveal that the vortices in the stratified fluid case are stretched along the streamwise direction in the near wake. The momentum thickness profiles show lower momentum thickness values for the stratified late wake compared to the non-stratified wake, implying that the drag on an accelerating hydrofoil in a stratified medium is reduced at the acceleration stage. The findings may improve our ability to predict drag due to maneuvering of micro-air/water-vehicles in stratified conditions.

## 1. Introduction

Stably stratified flows are common phenomena in the atmosphere and oceans and are more difficult for analytical, numerical and experimental investigation as they are typically characterized by variations (both in space and time) of the vorticity, energy and density or temperature. In non-stratified flows, instabilities are usually associated with the Reynolds number, $Re$. As the Reynolds number increases, inertial forces overcome viscous dissipation, and instabilities grow until they overtake the flow. In a stratified flow, a gravity stabilizing force is present, which generates buoyancy forces. Thus, for these kinds of flows one also needs to take into account the Froude number, $Fr$, in addition to $Re$ (Turner, 1973).

The ambient stratification at the pycnocline layer in the ocean has a unique effect on a vehicle wake through the formation of distinct long-time, large aspect ratio quasi-horizontal vortices (commonly termed as 'pancakes') and the radiation of internal waves from both the vehicle itself ('lee waves') or from its wake, as observed by Lin and Pao (1979). Thus, the stratified wake exhibits a distinct late-time signature that may potentially be traced directly to the generating body and used for identification of undersea vehicle. Monroe and Mei (1968) observed that the near-field wake growth in a turbulent stratified fluid is similar to that of a homogeneous fluid, but as the wake height grows above a critical value, further growth was inhibited.

---

[*] Corresponding author. Tel.: +972 0508782490. e-mail address: bengida@tx.technion.ac.il (H. Ben-Gida).

Browand et al. (1987) and Hopfinger (1987) reported $Nt$=1-2 (where $N$ is the Brunt-Väisälä frequency) as being the time at which wake turbulence first experiences the effect of the ambient stratification. In addition, Lin and Pao (1979) showed the occurrence of the first maximum wake height of a self-propelled slender body at a constant value of $Nt$=1.4 for 20<$Fr$<180, and at $Nt$=4.7 for towed slender bodies for 10<$Fr$<110. Hopfinger et al. (1991) reported that wake growth for a towed sphere at moderate Froude number (Fr≥4) was initially affected by stratification at around $Nt$≈2. Furthermore,

Spedding et al. (1996a-b) used particle image velocimetry (PIV) for quantitative measurements of the stratified wake velocity field behind a bluff body at long times ($Nt$>20). The authors characterized the evolution of the vortex wake of a towed sphere in a series of experiments with an independent variation of $10^3$<$Re$<$10^4$ and 1<$Fr$<10. Spedding et al. (1996b) found that the stably stratified wake width grows at approximately the same rate as in a three-dimensional non-stratified wake, but it becomes narrower, not wider, with decreasing $Fr$ (i.e., as stratification effects become stronger). The centerline defect velocity, on the other hand, reaches values an order of magnitude above those measured for three-dimensional non-stratified wakes at equivalent downstream locations.

Recently, Meunier (2012) performed dye visualization of the flow behind a vertically towed cylinder at 100<$Re$<200 and 0.8<$Fr$<1.8. It was found that stratification smoothed the von-Kármán vortices in the wake only when $Fr$≈1. When the Froude number was smaller than one, the von-Kármán vortices were formed again with higher wavelength.

The majority of experimental studies of stratified wakes reported in the literature used towing tanks. As a consequence, the first part of the wake corresponds to a transient stage; the steady state regime is achieved after a streamwise distance of the order of 20 body lengths in most cases, from which the wake data are extracted. Thus, on the one hand studies focused mainly on late time effects of the stratification on the wake, where the bluff body moves with constant velocity, while on the other hand, early stages of wake formations have been studied extensively (Huang et al, 2001; Sengupta et al., 2007, Freymuth et al., 1985; Dickinson and Götz, 1993; Jeon and Gharib, 2004), yet, not in a stratified environment.

Jeon and Gharib (2004) noted that for a constantly accelerating cylinder at 1,000<$Re$<2,000, at early times, the wake is symmetrical and the vortices in the wake do not interact strongly. The wake asymmetry increases at the time the body passed 4-6 diameters. This time depends on the acceleration. Dickinson and Götz (1993) investigated a wake of an accelerating thin wing model at Re=200 and several angles of attack, towed in a tank. The authors found that the flat wings generate a von-Kármán street for at least the first 7.5 chord lengths of travel. They showed that the oscillations of the forces following the inertial transients are similar to the discrete vortex theory predictions for a flat plate (Sarpkaya, 1975) and a conventional airfoil (Katz, 1981) as well as for flow patterns at a $Re$ numbers of 1,050 (Kiya and Arie, 1977).

Following Dickinson and Götz (1993) methodology, Huang et al. (2001) used PIV to study the flow around an accelerating wing model (NACA 0012; $c$=6 cm) at $Re$=1,200 and several angle of attacks towed in a water tank. The flow at low angles of attack presented stable vortex shedding in the wake, which was established after the initial period of complex vortex evolution on the wing surface. The goal of the aforementioned wing studies was to investigate how the acceleration from rest affects the aerodynamic parameters of the wing. The accelerations used were relatively high and the wing reached a steady state after about one-chord length of travel.

Summarizing the literature review we can conclude that, to the best of our knowledge, the case of stratified wakes behind accelerating bluff bodies from rest is not adequately addressed. In this study a hydrofoil with a NACA 0015 cross section was accelerated from rest at a constant rate inside a water tank. Throughout its acceleration the hydrofoil was subjected to stable stratification with low Reynolds and Froude numbers. Velocity fields were extracted in the wake region using a PIV technique (Section 2). Characterization of both non-stratified and stratified wakes is achieved through the calculation of the vorticity, circulation and momentum thickness (Section 3). Section 4 is devoted to summary and conclusions.

## 2. Experimental Setup

Measurements of the wake downstream of an accelerating towed hydrofoil were taken in a tank using a PIV system (see Fig. 1).

### 2.1. The tank facility

The tank is composed of glass and it is open to air from above with dimensions of 50 x 20 x 20 cm$^3$ (length x height x width). The streamwise, vertical and spanwise directions are defined by *x*, *y*, and *z* respectively. A linear displacement mechanism, powered by an external air pressure (3 atm) acting on an inside piston, is placed on top of the tank to allow the movement of the hydrofoil. The displacement mechanism has a 10 cm length slide plate on which the hydrofoil is mounted. The hydrofoil velocity has been controlled by the pneumatic system.

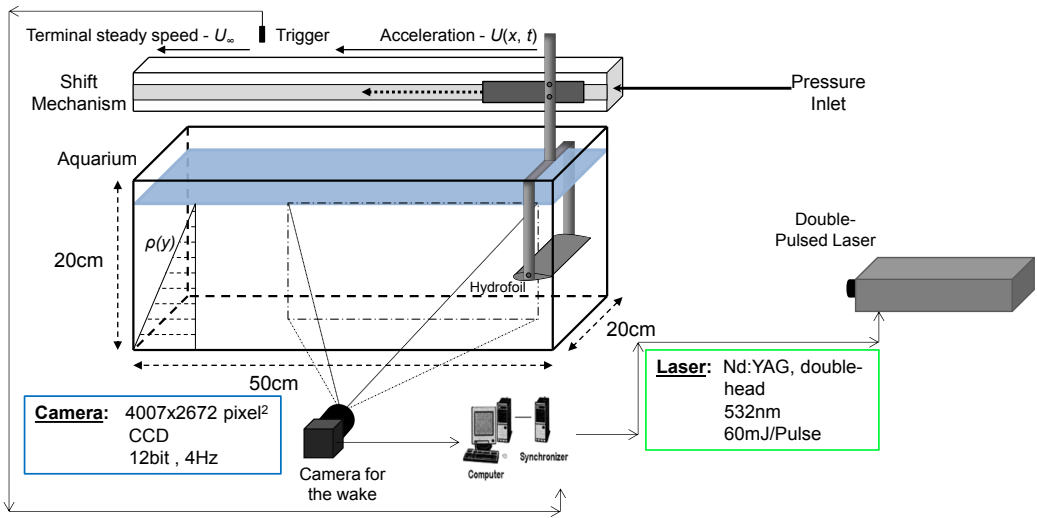

**Fig. 1.** Experimental setup at rest with stratification conditions (same as for water).

### 2.2. Hydrofoil

The cross-section of the hydrofoil is a NACA 0015, as it accounts for one of the simplest cases to study: wing with a symmetrical cross-section, which does not generate lift at zero angle of attack. The chord length, *c*, is 5 cm and the wingspan, *b*, is 18 cm, designed to reduce three-dimensional effects originating from the wingtips by the walls of the tank.

## 2.3. PIV system

The PIV system consists of a double-head pulsed Nd:YAG laser (nano-PIV, Litron, UK), emitting energy of 60 mJ/pulse with a wavelength of 532 nm. A thin laser sheet (2 mm wide) was set to illuminate the wake behind the mid-plane of the hydrofoil in the streamwise-vertical ($xy$) plane (see Fig. 2). The flow was seeded (both in water and stratified fluids) with hollow glass spheres (110P8, Potters Industries) with an average diameter of 10 μm.

The PIV camera was an 11 megapixel (4008 x 2672 pixel$^2$) double exposure CCD camera with a 12 bit dynamic range equipped with an AF micro-nikkor 60 mm lens that was positioned perpendicularly to the laser sheet. The frame rate used by the camera, synced with the lasers was 2 Hz with the time interval between two laser pulses adjusted from the PIV software (Insight 3G, TSI Inc.) by the synchronizer (TSI 610035). The CCD camera field of view was approximately 35 x 23 cm$^2$ (length x height) or $7c$ by $4.6c$. Therefore, the resolution was 88 μm/pixel. The capture of the PIV images was set to record image pairs only when the hydrofoil crosses the infrared trigger device (see Fig. 2).

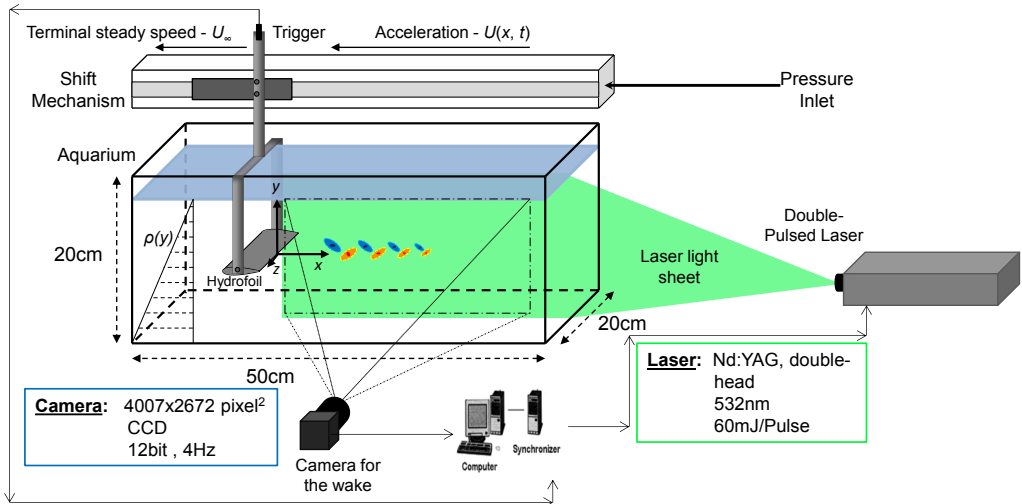

**Fig. 2.** Experimental setup when PIV images are taken with stratification conditions (same as for water).

## 2.4. Procedure

The experiments were carried out by accelerating a NACA 0015 hydrofoil from rest at a constant rate ($a$=1.5 cm/sec$^2$), subjected to strong stratification and low Reynolds number, from one side of the tank to another. At some distance, the hydrofoil crossed an infrared trigger device that signaled the PIV system to record image pairs of the hydrofoil wake.

The glass tank was filled separately with two fluids: fresh water and a mixture of different saline solutions that created a stable stratified layer with a vertical density gradient of $d\rho/dy$=129 kg/m$^4$ according to the modified Oster method (Zurigat et al., 1990). An oscillation-type density meter (DA-130N KEM$^{TM}$; resolution of 0.1 kg/m$^3$) was used to measure the density in the tank. The hydrofoil wake was characterized for both stratified and non-stratified flows, separately, for comparison.

When investigating stratified flows through PIV measurements it is essential that the fluid illuminated by the laser will have a constant refractive index. In this study, the most concentrated saline solution had a concentration of 29 g/L with a refractive index of 1.3379 (Daviero, 2001). Therefore, the maximum variation of the refractive index of the stratified fluid, compare to the fresh water (=1.3329), is 0.4%. Consequently, we can state that the refractive index of the fluid inside the tank was nearly constant and had no effect on the PIV images.

The acceleration period for the hydrofoil ended after a run distance of 6.2$c$, from which the hydrofoil had a terminal steady velocity of $U_\infty$=10 cm/sec. Both the stratified and the fresh water flows had a terminal steady Reynolds number ($Re_\infty$) of $\rho_0 U_\infty c/\mu_0$=5,000, where $\rho_0$ and $\mu_0$ are the average density and dynamic viscosity of the fluid in the tank, respectively. In the stratified experiments, the terminal steady Froude number ($Fr_\infty$) is $U_\infty/cN$=2, where $N$ is the Brunt-Väisälä frequency (=1.1 sec$^{-1}$) defined as $N^2 = (-g/\rho_0)d\rho/dy$.

Two triggering times of the PIV system were set, such that the hydrofoil wake was recorded once the hydrofoil moved about 22 cm corresponding to 3.2 sec, (we denote this state as early triggering/wake time) and after 38 cm or 5.1 sec (denoted hereinafter as late triggering/wake time). The PIV data of the early wake correspond to the hydrofoil during its acceleration stage with the velocity of $U_{80}$=0.8$U_\infty$ (=8 cm/sec), while the PIV images of the late wake captured the hydrofoil wake at the end of the acceleration phase with the terminal velocity $U_\infty$ (=10 cm/sec). Thus, early wake results correspond to $Re_{80}$=4,000 and $Fr_{80}$=1.6, whereas late wake results correspond to $Re_\infty$=5,000 and $Fr_\infty$=2.

Four sets of experiments are presented; two sets of stratified and non-stratified experiments for a run distance of 22 cm and another two sets of stratified and non-stratified experiments for a run distance of 38 cm. Each set of experiments contained an average of 50 runs of the hydrofoil along the tank, providing sufficient statistics for the ensemble averaging and the following analysis of fluctuations. In each run a pair of PIV images was taken (see Fig. 2) to extract the related velocity map. During the non-stratified flow experiments, the hydrofoil was immediately returned to its initial position (right side of the tank) after each run, following a settling time of three minutes to allow the flow in the tank to decay. During the stratified flow experiments, the hydrofoil was returned to its initial position two minutes after each run, following a settling time of five minutes to allow the flow to decay and the stratified layer to resettle in the tank. We validated that the stratified layer resettled to its initial state before the start of each run by measuring the density gradient in the water tank using the density meter.

*2.5. Data processing*

The flow field behind the hydrofoil was calculated using interrogation windows of 32 x 32 pixel$^2$ with 50% overlap between neighboring spots and applying a FFT cross-correlation function on each interrogation area and sub-pixel Gaussian interpolation, yielding a spatial resolution of 18 vectors per chord. PIV analysis was performed using open source PIV software, OpenPIV (www.openpiv.net, see Taylor et al., 2010)

Two distinct areas behind the hydrofoil are defined, each for a different triggering time. Fig. 3-Fig. 4 show the camera field of view during the experiments with the hydrofoil location at rest and when the PIV images were taken. For the early triggering/wake time the PIV analysis area is approximately 14 x 10 cm$^2$ (length x height) or 2.8$c$ by 2$c$, and for the late triggering/wake time it is 27 x 10 cm$^2$ (length x height) or 5.4$c$ by 2$c$.

The coordinate system was chosen in a way that the $x$-direction starts from the trailing-edge of the hydrofoil and continues right into the wake, the $y$-direction is pointing upward and the spanwise direction is deduced according to a right-hand coordinate system (see Fig. 1).

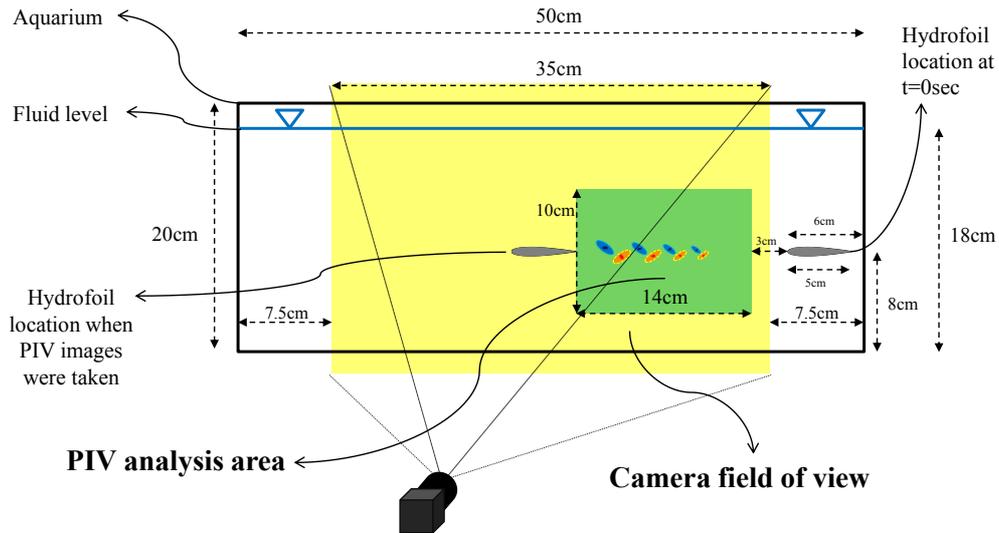

**Fig. 3.** PIV analysis area (in green) for the early triggering/wake experiments.

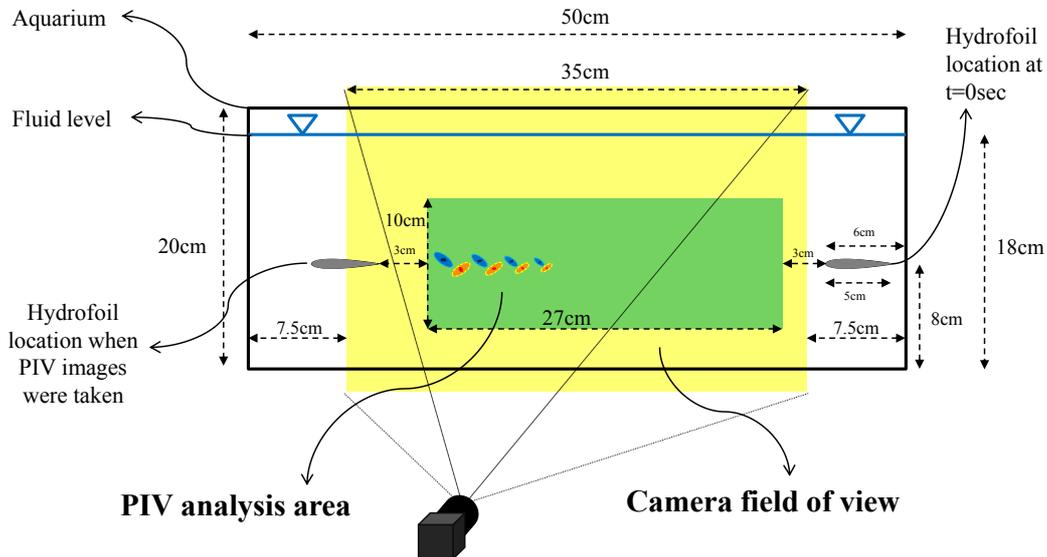

**Fig. 4.** PIV analysis area (in green) for the late triggering/wake experiments.

## 3. Results and Discussion

The following sections describe the boundary conditions set for the experiments and the observed effect of stratification on the wake of the hydrofoil. Comparison between the stratified and non-stratified flows is made in terms of vorticity maps, circulation distribution throughout the wake and momentum thickness.

An error analysis based on the root sum of squares method has been applied to the velocity data. The errors were estimated as: 2.5% for the velocity values and 10% for the vorticity (Raffel et al., 2007).

### 3.1. Boundary conditions

When accelerating a hydrofoil through a stratified medium in a tank one may need to account for several effects that can have a direct influence on the evolving wake. In the current study we investigated two main effects, which can have a non-negligible impact on the wake of the hydrofoil; these are the zero-frequency waves and the added mass effects.

According to Xu et al. (1995) any moving disturbance (e.g., hydrofoil) in a stratified fluid may produce horizontal jet-like columnar disturbances of zero-frequency propagating upstream of the source. These disturbances propagate far ahead of the source, interact with the end walls of the tank, and can reflect back and interact with the motion field of the hydrofoil. The position of the fastest first mode disturbance front, with respect to the source, was estimated according to Xu et al. (1995) and found to be located at 0.38 m in the early wake PIV images and at 0.49 m for the late wakes. The terms 'early' and 'late' refer herein to time domain. Hence, we can state that for both early and late wakes the disturbance front position was less than 0.50 m, which is the tank wall position (see Fig. 1); i.e., it is shown that when the PIV images were captured, the disturbance front did not reach the tank wall, thus, the hydrofoil wakes shown in the current study are presumably not affected by zero-frequency waves.

It is known that an accelerating body in a viscous fluid medium also accelerates the fluid around it. As a result, work must be done to increase the kinetic energy of the fluid around it as well as to increase the kinetic energy of the body. Daniel (1984) termed the force required to increase the kinetic energy as the acceleration reaction, which acts in addition to drag in order to resist motion. We define the acceleration reaction coefficient as follows:

$$C_{ar} = \frac{2\alpha V}{U_\infty^2 S_{ref}} a \qquad (1)$$

where $\alpha$ is the added-mass coefficient estimated as 0.1 for the NACA 0015 hydrofoil, according Daniel's (1984) theoretical analysis for elliptical cylinder with $t/c$=0.15. $V$ is the volume of the accelerated hydrofoil (=3·10$^{-4}$ m$^3$), $a$ is its acceleration relative to the fluid (=1.5 cm/sec$^2$) and $S_{ref}$ is the reference area of the hydrofoil (=$b·c$=9·10$^{-3}$ m$^2$). The acceleration reaction coefficient was calculated for each set of experiments and found to be bound by a value of $C_{ar}$=1.5·10$^{-3}$. For comparison, experimental results reported by Sheldahl and Klimas (1981) showed that the drag coefficient of a NACA 0015 at $Re_\infty$=10,000 is $C_d$=3.6·10$^{-2}$. It should be noted however that for the Reynolds number used in the current study (5,000) the drag coefficient is expected to be higher. Consequently, we can claim that the added mass effect in the current study is negligible and with a maximum value of 4% of the drag force acting on the hydrofoil.

*3.2. The early wake*

In the following section, we characterize the early wake in terms of velocity fluctuations and vorticity content in order to depict the flow structures form in both non-stratified and stratified fluids. The spanwise vorticity in the wake is computed directly from the PIV data using a least squares differentiation scheme (Raffel et al., 2007). The values of the instantaneous spanwise vorticity, $\omega_z(x,y)$, were averaged for each set of experiments, to evaluate an ensemble average of the spanwise vorticity maps, $\Omega_z(x,y)$. The resultant ensemble average maps were multiplied by the hydrofoil chord, $c$, and divided by the 80% velocity of the hydrofoil, $U_{80}$, to deduce normalized ensemble average vorticity ($\Omega_z·c/U_{80}$) maps.

Fig. 5 shows contour maps of the normalized ensemble average of spanwise vorticity in the early wake behind the hydrofoil accelerated through a non-stratified and a stratified medium, where the horizontal and vertical axes are normalized by the hydrofoil chord, $c$. The range of the normalized ensemble average vorticity in each image is -3 to 3. In addition, velocity vectors, calculated from the PIV, depict the flow patterns in the wakes. Since the hydrofoil was accelerated inside the tank of initially stagnant fluid, the velocity vectors depicted in Fig. 5 are represent the velocity filed as observed behind the hydrofoil moving to the left.

Moreover, an additional horizontal time axis, *t*, was added to Fig. 5, where *t* is defined as the 'wake age'; i.e., the time, *t*, originates from the trailing-edge of the hydrofoil at its final position, where the wake is created, and increases throughout the wake pattern, where the wake is already developed and 'older' as it was created *t* seconds ago when the hydrofoil trailing-edge passed that point. The constant acceleration of the hydrofoil, *a*, and the respective values of the hydrofoil velocity distribution, *U(x,t)*, were used to establish a relation between the space vector, *x*, and the time, *t*. Using this parabolic relation, each *x* location throughout the wake was transformed into the time space, *t*. For simplicity, the early wakes behind the accelerated hydrofoil depicted in Fig. 5 are plotted with the normalized time scale, $at/U_{80}$.

*3.2.1. Stratification effects*

As it could be expected, in both the non-stratified and stratified fluids, the upper part of the wake is composed of negative spanwise vorticity (clockwise rotation) and the bottom part of the wake by positive spanwise vorticity (counter-clockwise rotation). Our first observation from the early wake vorticity maps (see Fig. 5) relates to the similar normalized vorticity values forming in both the non-stratified (Fig. 5a) and stratified (Fig. 6b) fluids; i.e., there is no immediate evidence of the stratified fluid affecting the vorticity distribution behind the accelerated hydrofoil, especially in the near wake region (below $x/c \approx 1.0$).

Yet, a closer examination of the velocity vectors in Fig. 5b shows a distinctive wake pattern in the region of $x/c > 1.0$. In contrast to the non-stratified wake, where the velocity vectors show a typical wake flow pattern behind a moving hydrofoil (velocity vectors in the wake are directed to the left), the stratified wake comprised of velocity vectors that are due to the stratified fluid effect in the wake region of $at/U_{80} > 0.15$ ($t > 0.8$ sec). One may notice the velocity vectors above and below the wake core ($y/c=0$), in the region of $at/U_{80} > 0.3$ ($t > 1.6$ sec), that could lead to a vortical flow in the far wake (at early times of acceleration), which is absent in the non-stratified wake.

As mentioned in section 2.4, the stratified layer in the current study is characterized with a Brunt-Väisälä frequency of $N=1.1$ sec$^{-1}$; i.e., the effect of the stratified layer on the wake is time dependent with a time period of $T=0.9$ sec. Accordingly, the wake region at $at/U_{80} > 0.15$ ($t > 0.8$ sec) is showing an image of a wake being manipulated by the stratified layer from one to slightly more than two time periods. Apparently, the 3.2 sec of constant acceleration (corresponding to the early wakes), is the sufficient time span for the stratified fluid to respond and deform the wake of the hydrofoil, as it appears in Fig. 5b.

In the next section we characterize the effect through comparison of the non-stratified and stratified late wakes, depicted after a longer acceleration time of 5.1 seconds.

*3.3. The late wake*

The normalized ensemble average vorticity ($\Omega_z \cdot c/U_\infty$) maps in the late wake were computed in the same manner as described in section 3.2, except that $U_\infty$ is used as a normalization factor instead of $U_{80}$. Fig. 6 shows contour maps of the normalized ensemble average of spanwise vorticity in the late wake behind the hydrofoil accelerated through the non-stratified and stratified fluid. The hydrofoil chord was used to normalize both horizontal and vertical axes. The velocity vector fields, as observed from the accelerated hydrofoil, were calculated from the PIV realizations in order to emphasize the flow patterns in the wake. The range of the normalized ensemble average vorticity in each image is between -3 to 3. This means that compared to the early wake values, the average vorticity amplitude is about 25% higher. The same type of the normalized horizontal time axis, $at/U_\infty$, was added to Fig. 6, to emphasize the fact that the wake is more developed because of the time it took for the hydrofoil to pass from the right to the left side of the figure.

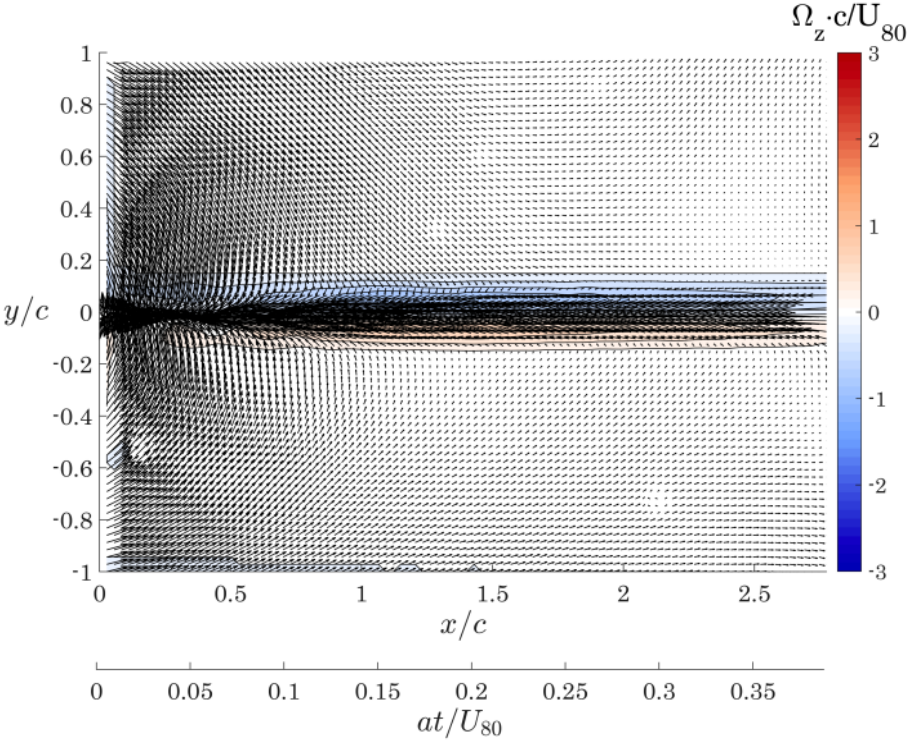

(a)

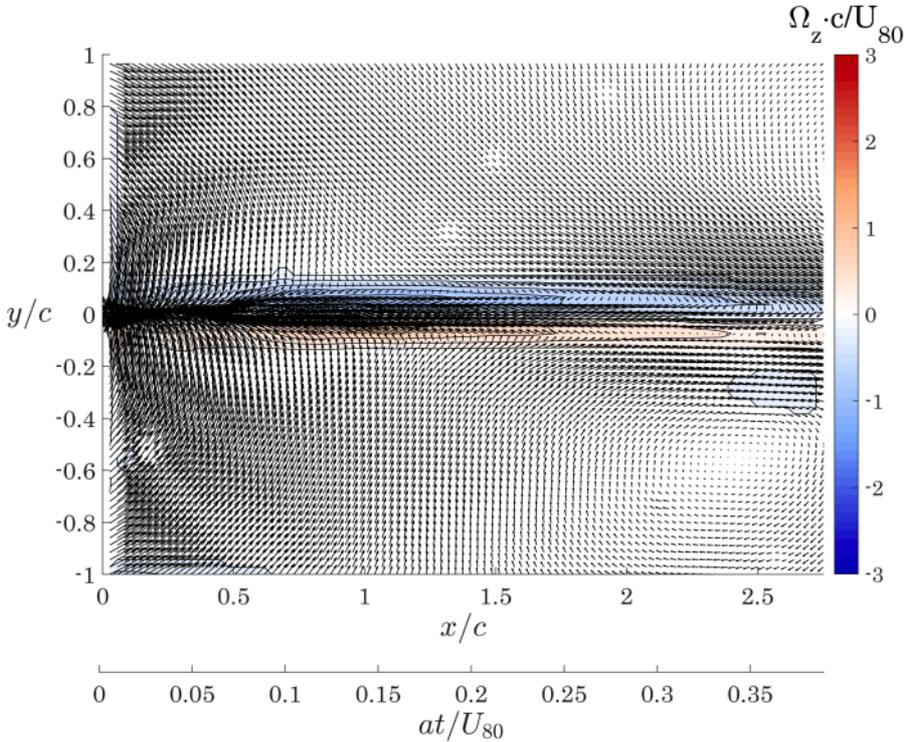

(b)

**Fig. 5.** Velocity vectors and contours of the normalized ensemble average spanwise vorticity, $\Omega_z \cdot c/U_{80}$, of the early wake behind a hydrofoil accelerating through fresh water ($Re_{80}$=4,000) and a stratified layer ($Re_{80}$=4,000, $Fr_{80}$=1.6) for 3.2 sec. (a) Early wake evolution in fresh water; (b) early wake evolution in a stratified layer.

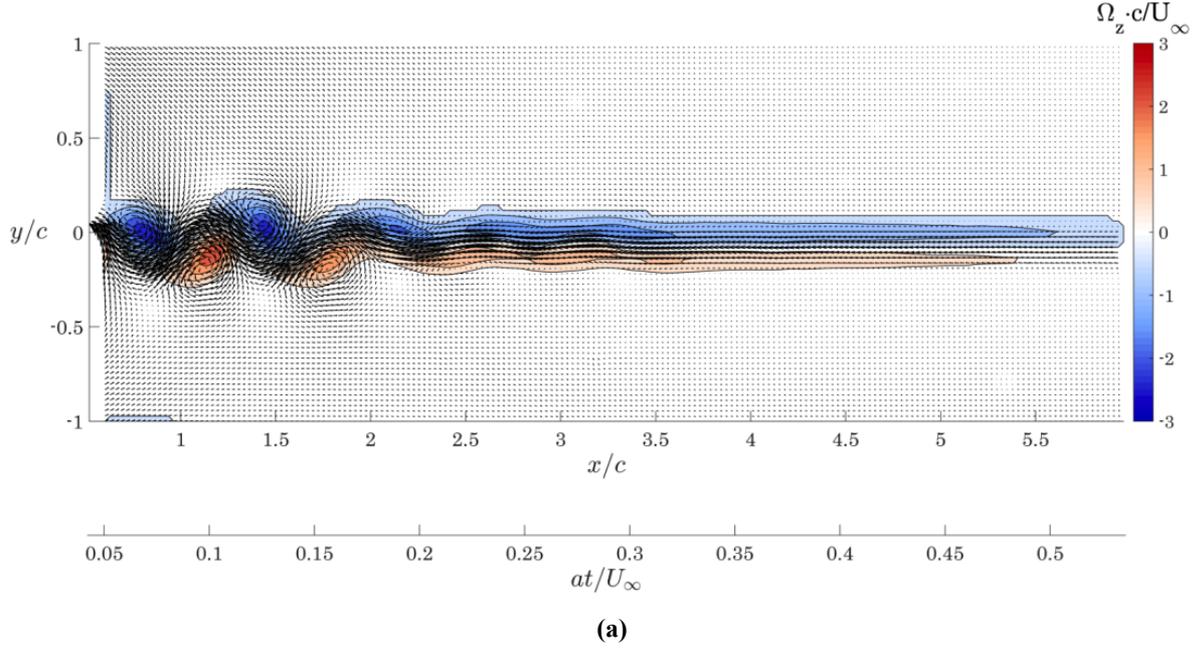

(a)

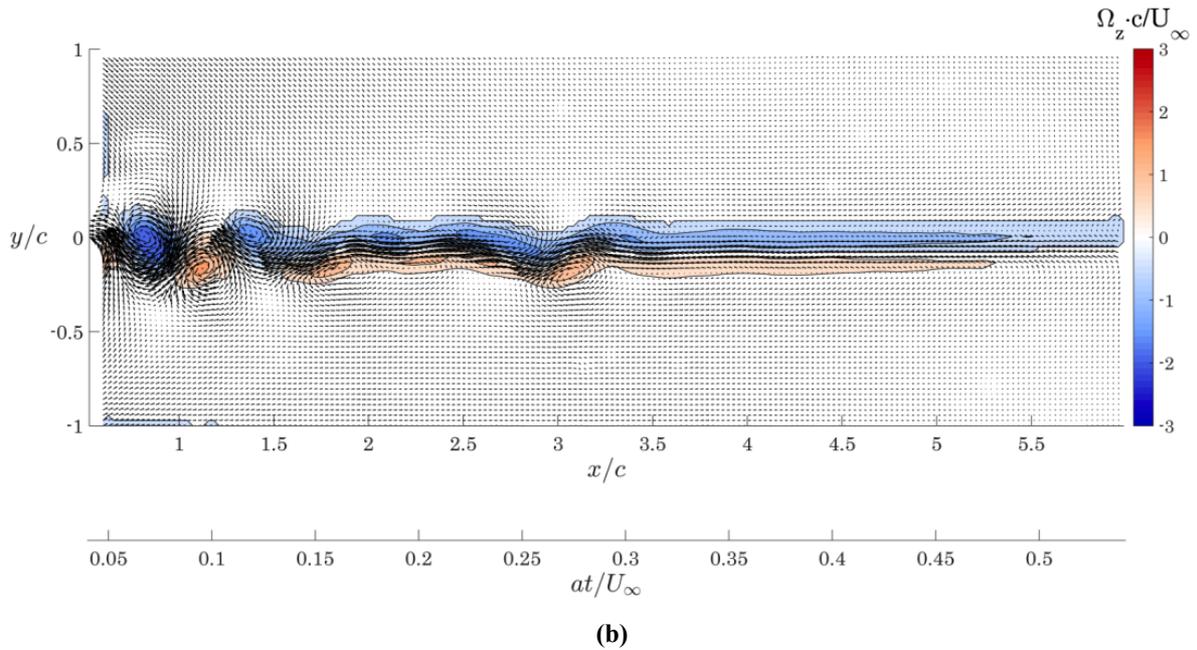

(b)

**Fig. 6.** Velocity vector fields and contours of the normalized ensemble average spanwise vorticity, $\Omega_z \cdot c/U_\infty$, of the late wake behind a hydrofoil accelerating through fresh water ($Re_\infty$=5,000) and a stratified layer ($Re_\infty$=5,000, $Fr_\infty$=2) for 5.1 sec. (a) Late wake evolution in fresh water; (b) late wake evolution in a stratified layer.

*3.3.1. Stratification effects*

One can observe that Fig. 6a, describing the wake evolution through a non-stratified medium, shows a considerably different wake pattern than Fig. 6b, describing the wake evolution through a stratified layer ($Fr_\infty$=2). The stratified layer effect on the late wake of the hydrofoil at low Reynolds number ($Re_\infty$=5,000) is manifested through weaker vortices that are also more stretched in the horizontal direction, shown by lower normalized vorticity values compare to those in the non-stratified wake. These observations are in agreement with previous results obtained in different settings (e.g., Chomaz et al., 1993; Spedding, 1997), thus confirming that the effect of the stratified fluid on the wake of accelerating bluff body in the given range of values is also characterized by vortices that are being stretched in the horizontal direction and thinner in vertical direction.

Here, the stratification effect can be seen continuously throughout the entire late wake evolution, in contrast to the early wake evolution in the stratified layer (see Fig. 5b), where the stratification was shown to affect the wake in the region of $at/U_\infty$>0.15. The continuous effect of the stratified layer on the late wake may arise due to the distinctive wake pattern characterized by vortices being shed into the wake. The following sub-section contains comprehensive characterization of the late wake pattern.

*3.3.2. The low $Re_\infty$ effect*

There have been many studies over the years on the nature of low Reynolds flows, in the range of $10^3$<$Re$<$10^5$, over wings and airfoils. Few examples are Jacobs and Sherman (1937), Neuhart and Pendergraft (1988), Huang and Lin (1995) and Yarusevych et al. (2009). All concluded that if large velocity gradients exist over an airfoil then the two shear layers shedding from the trailing-edge mix together, generating an oscillatory organized vortex street. In the current study the hydrofoil accelerate from rest and thus the flow field about the hydrofoil is characterized by large velocity gradients. Indeed, one can observe an apparent organized vortex street structure throughout the acceleration of the hydrofoil in the non-stratified medium, as depicted in Fig. 6a. Yet, such organized vortex street pattern is seems disturbed in the stratified case (see Fig. 6b), which has an additional buoyancy force.

Analyzing the organized vortex street structure in the non-stratified wake, we can observe that it starts at $x/c$≈2.5 (see Fig. 6a). Thereby, the hydrofoil has a certain Reynolds number ($Re_c$) from which an organized vortex street is starting to form in its wake. During the late wake experimental setup, the hydrofoil was accelerated at a constant rate along the tank for a distance of 31 cm ($x/c$=1.3 in Fig. 6a) until it reached a constant terminal velocity $U_\infty$. From the linear velocity profile of the accelerated hydrofoil along the tank we can deduce the local Reynolds number of the flow at $x/c$≈2.5; that is about 3,800. As already mentioned, the early wake images (see Fig. 5) were captured when the hydrofoil reached a velocity corresponding to $Re_{80}$=4,000. An organized vortex street was absent in the early wake images presumably due to the time lag that takes vortices to shed into the wake region. In the following sub-section we identify the vortices in the stratified and non-stratified wakes and study their evolution throughout the wake.

*3.3.3. Vortex identification and circulation of the wake*

Characterization of the vortex structures in the late wake can be done using a vortex identification algorithm, which involves a critical-point analysis of the local velocity gradient tensor and its corresponding eigenvalues (Chong et al., 1990; Adrian et al., 2000). For two-dimensional flow the determinant of the velocity tensor will either have two real eigenvalues ($\lambda_{r1}$, $\lambda_{r2}$) or a pair of complex conjugate eigenvalues ($\lambda_{cr} \pm i\lambda_{ci}$). Adrian et al. (2000) noted that the strength of any local swirling motion is quantified by $\lambda_{ci}$, which is the swirling strength of the vortex. Consequently, we can identify vortex regions in the late wake by finding their cores, which are the local peak values of the swirl function defined as:

$$\lambda_{ci} = \text{Im}\left[\sqrt{\frac{1}{4}\left(\frac{\partial u}{\partial x}+\frac{\partial v}{\partial y}\right)^2 + \frac{\partial u}{\partial y}\frac{\partial v}{\partial x} - \frac{\partial u}{\partial x}\frac{\partial v}{\partial y}}\right] > 0 \qquad (2)$$

Fig. 7 shows contour maps of the swirl function, $\lambda_{ci}$, (vortex regions) in the late wake behind the hydrofoil accelerated through a non-stratified and a stratified medium. The vortex cores, defined by the local swirl function maximum $\lambda_{ci\text{-}max}$, are marked by crosses in Fig. 7.

An additional normalized horizontal time axis, $at/U_\infty$, was added to Fig. 7, computed in a similar manner to the late wake analysis.

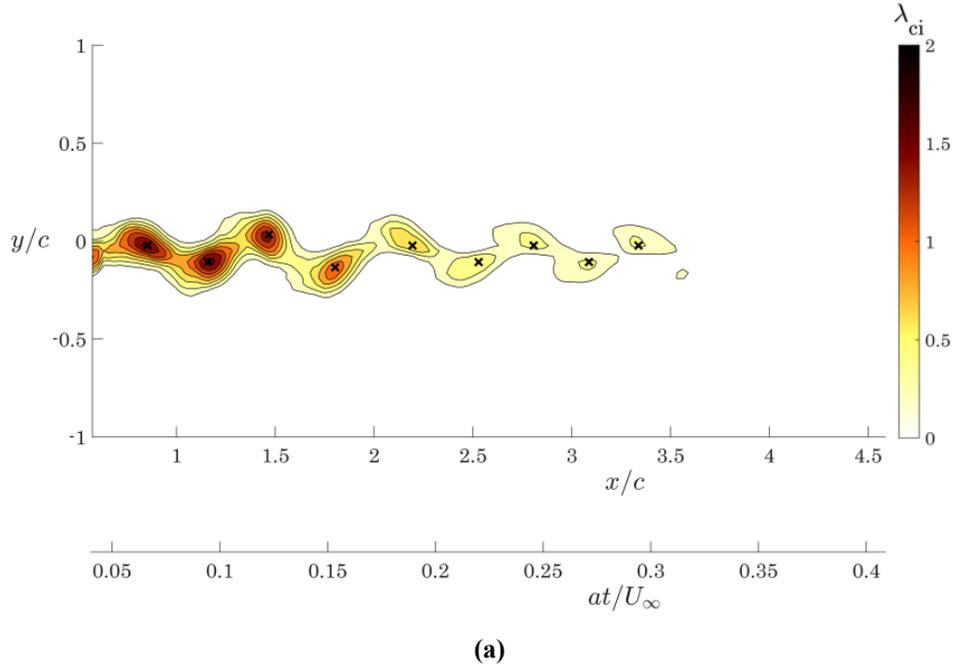

(a)

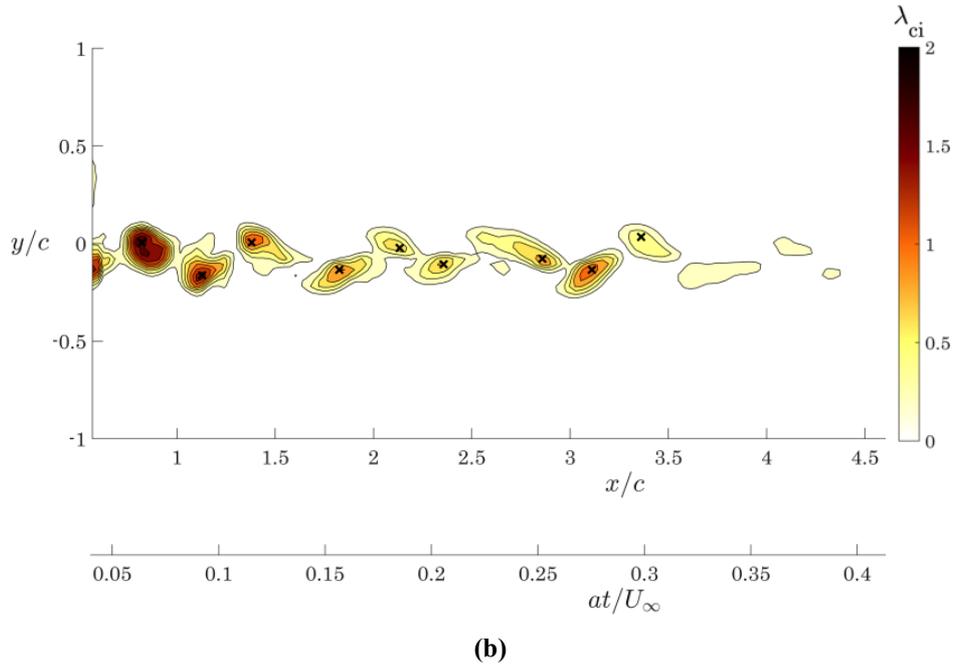

(b)

**Fig. 7.** Swirl function contours of the late wake behind a hydrofoil accelerating through: (a) fresh water ($Re_\infty$=5,000) and (b) a stratified layer ($Re_\infty$=5,000, $Fr_\infty$=2). The 'x' symbols represent the vortex cores.

The non-stratified late wake (Fig. 7a) shows intense vortex regions that are characterized by a decrease of the swirl function (vortex strength) throughout the wake, from $x/c$=0.5 to $x/c$≈2.5. We recall that $x/c$≈2.5 was found to be the location from which the organized vortex street is starting to form (see sub-section 3.3.2). However, when introducing the wake into a stratified medium (Fig. 7b) one can observe weaker vortex regions that are being developed in the wake for longer time, up to $at/U_\infty$≈0.4. In comparison, the vortex regions in the non-stratified wake were developed up to $at/U_\infty$≈0.3. We may postulate that the stratified wake signature is weaker, sustains longer in time and exhibits lower dissipation rates than the non-stratified wake.

Characterizing the late wake pattern in terms of the circulation may add more insight on such distinctive vortex distribution throughout the stratified medium. The derivation of the circulation distribution throughout the wake is based on the vortex identification analysis. The area of each vortex in the wake was defined as the area in which the swirl function values are at least 50% of the peak swirl value in the vortex core ($\lambda_{ci} \geq 0.5 \cdot \lambda_{ci-max}$). The vortex streets depicted in Fig. 6 were characterized in terms of circulation distribution by integrating the ensemble vorticity values surrounded by the outer boundary of each positive- and negative-signed vortex. The circulation of each positive- and negative-signed vortex with an ensemble vorticity distribution $\Omega_z(x,y)$ was calculated numerically throughout the vortex area as follows:

$$\gamma^{+/-} = \iint \Omega_z(x,y) dx dy \qquad (3)$$

Eq. (3) indicates that counter-clockwise vortices (positive vorticity regions depicted in Fig. 6) cause positive circulation, which contributes to lift generation according to Kelvin's circulation theorem. In each vortex area, the minimum absolute ensemble vorticity value was identified and stored as $|\Omega_z|_{min}$. The total circulation of each vortex was then calculated by assuming a normal Gaussian distribution of the vorticity outside the vortex area ($|\Omega_z| < |\Omega_z|_{min}$):

$$\Gamma^{+/-} = \left(1 + \frac{|\Omega_z|_{min}}{|\Omega_z|_{max}}\right) \gamma^{+/-} \qquad (4)$$

where $|\Omega_z|_{max}$ is the maximum vorticity value within the vortex area (e.g., Spedding et al., 2003).

Fig. 8 depicts the positive and negative normalized circulation values, $\Gamma^{+/-}/cU_{\infty,avg}$, as well as the normalized absolute circulation, $|\Gamma|/cU_{\infty,avg}$, throughout the late wakes that evolve behind the hydrofoil when accelerating it through the non-stratified and the stratified fluid. It is shown that accelerating the hydrofoil throughout the tank causes the circulation in the late wake to increase for both the non-stratified and the stratified layer fluids. We can observe that the non-stratified wake is characterized by higher absolute circulation values than the stratified wake (see Fig. 8a). This qualitative observation is consistent with the vorticity and swirl function maps (see Fig. 6-Fig. 7), in which the stratification effect was seen continuously throughout the evolution of the late wake.

Moreover, Fig. 8b shows that the non-stratified medium allows the circulation to grow much faster (at larger $x/c$) throughout the wake than the stratified medium. The sharp increase of the circulation in the non-stratified wake starts around $x/c=2.2$, whereas the circulation in the stratified wake undergoes similar increase only around $x/c=1.4$. Apparently, the wake region from $x/c=1.4$ and further downstream the wake is strongly affected by the stratified medium, which causes lower circulation values compare to those observed in the non-stratified medium. The following section discusses the momentum thickness distribution throughout the non-stratified and stratified wakes in order to acquire more quantitative insight regarding the stratification effect observed in previous sections.

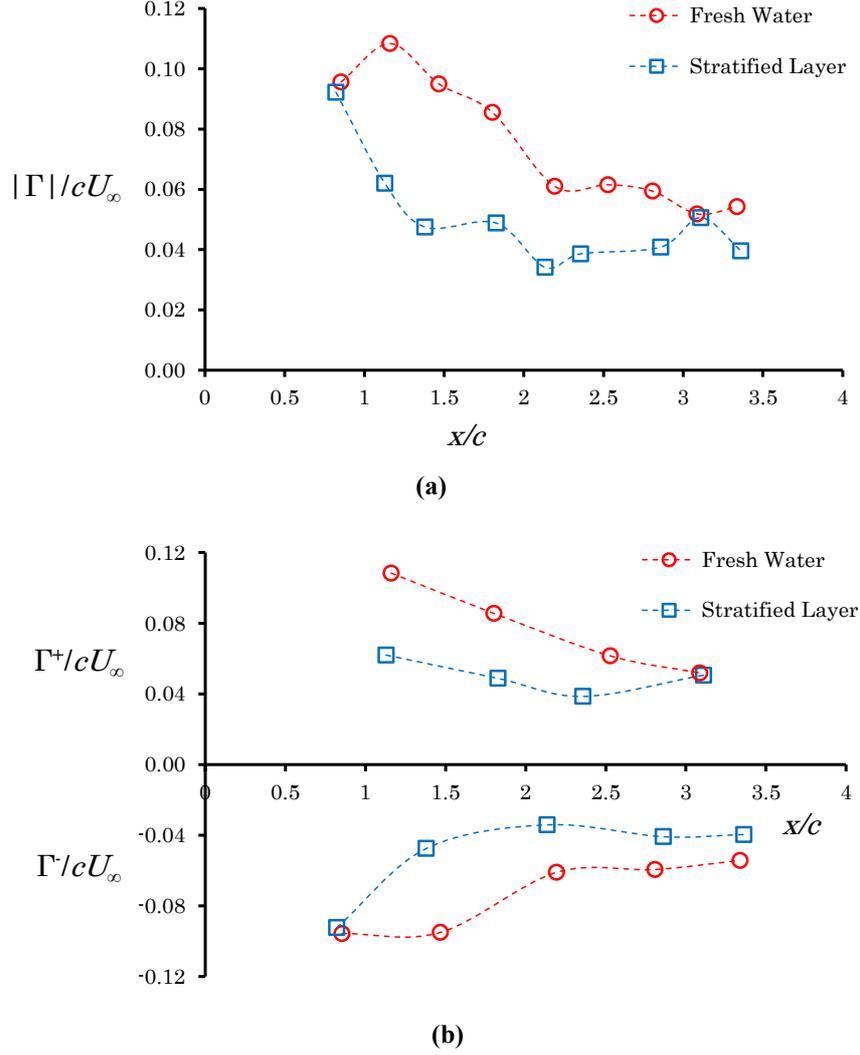

**Fig. 8.** (a) Normalized circulation and (b) normalized absolute circulation distributions throughout the late wake that evolves in fresh water ($Re_\infty$=5,000) and a stratified layer ($Re_\infty$=5,000, $Fr_\infty$=2). Each mark indicates the circulation value of a single positive- or a negative-signed vortex.

*3.4. Momentum thickness*

    The previous section has shown that the stratified layer has a strong effect on the late wake characteristics behind the accelerating hydrofoil; the vortex deformation and the circulation reduction are the two major examples. Yet, a more profound quantitative analysis of the early and late wakes is essential to better understand the stratification effect. Here, we calculate the momentum thickness distribution throughout the early and late wakes to provide such quantitative insight. We start our analysis by calculating the streamwise velocity profiles throughout the hydrofoil wake as it evolves through the stratified and non-stratified fluids.

    A comparison between the different velocity profiles $U(x,y,t)$ in the wake can be performed only if the accelerating hydrofoil is observed from a different point of view; as if the observer moves with the hydrofoil and accelerates with it. Transformation to a coordinate system attached to the accelerating hydrofoil was accomplished by de-trending the streamwise velocity vectors in each PIV instantaneous velocity map, *i*, using the hydrofoil velocity profile along the tank, $U(x,t)$, according to:

$$u'(x,y)_i = u(x,y)_i + U(x,t) \tag{5}$$

The resultant velocity vectors, $u'(x,y)$, were normalized by the hydrofoil velocity, $U(x,t)$, for each PIV velocity map, $i$. Then, the normalized instantaneous streamwise velocity profiles, $u'/U$, were ensemble averaged for each $x_s$-location, providing normalized ensemble average streamwise velocity profiles as follows:

$$\left(\frac{u'}{U}\right)_{avg,x_s} = \frac{1}{n}\sum_{i=1}^{n}\frac{u'_i(x_s,y)}{U_i(x_s)} \tag{6}$$

where $x_s$ is a specific $x$-location along the wake pattern. It should be pointed out that the de-trending process is linear since the hydrofoil was constantly accelerated throughout the tank.

The wake momentum thickness is a useful tool to estimate the amount of streamwise momentum lost from the free stream due to the presence of a bluff body. In other words the momentum thickness is proportional to the drag force produces by the bluff body. Thus, wakes with larger momentum thickness will correspond to higher drag on a bluff body. In the current study we calculate the normalized momentum thickness for the ensemble wakes as follows:

$$\theta = \frac{1}{c}\int_{y=-c}^{y=c}\left(\frac{u'}{U}\right)_{avg}\cdot\left[1-\left(\frac{u'}{U}\right)_{avg}\right]dy \tag{7}$$

where the chord length of the hydrofoil was used as a normalization factor. The integration presented in Eq. (7) has been performed in the vertical direction, from $y=-c$ to $y=c$, for each $x$-location throughout the ensemble (early and late) wakes of the stratified and non-stratified fluids. As a result, a distribution of the momentum thickness has been acquired for each set of experiments.

Fig. 9 depicts the distribution of the normalized momentum thickness throughout the stratified and non-stratified wakes in the early and late triggering times.

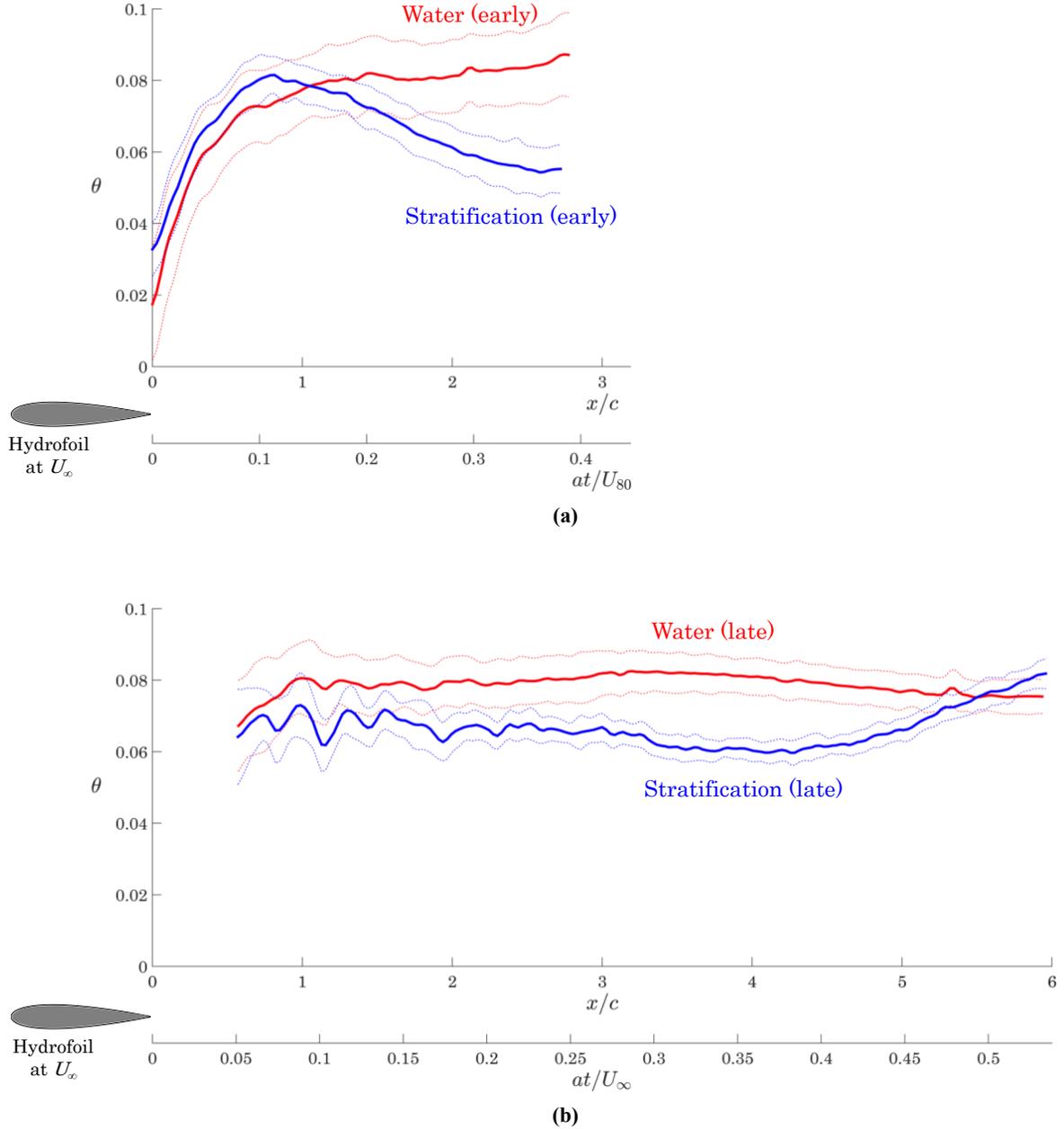

**Fig. 9.** Distribution of the normalized momentum thickness (solid lines) throughout: (a) early wakes evolved behind a hydrofoil accelerating through fresh water ($Re_{80}$=4,000; red) and a stratified layer ($Re_{80}$=4,000; $Fr_{80}$=1.6; blue); (b) late wakes evolved behind a hydrofoil accelerating through fresh water ($Re_{\infty}$=5,000; red) and stratified layer ($Re_{\infty}$=5,000; $Fr_{\infty}$=2; blue). The normalized momentum thickness curves are bounded on either side by the standard deviation (dashed lines).

The effect of stratification on the hydrofoil wake, as described in prior sections, is further supported by the quantitative momentum thickness analysis (see Fig. 9), where the reduction of the momentum thickness is related to vortex stretching and thinning. Early wake results (see Fig. 9a) show lower momentum thickness values in the stratified medium compare to the non-stratified medium in the region of $at/U_{80}$>0.15. This observation is consistent with the one in sub-section 3.2.1, where the velocity vectors were seen as being deformed in the stratified layer (see Fig. 5b). In addition, while the momentum thickness in the non-stratified medium increases with $x/c$ throughout the entire wake length, the momentum thickness in the stratified medium increases with $x/c$ (somewhat similar to the non-stratified case) until a distance of $x/c$≈1 ($at/U_{80}$≈0.15), from which it decreases.

Late wake results (see Fig. 9b) show lower momentum thickness values in the stratified medium compare to the non-stratified medium throughout the entire wake length, implying the stratification effect on the wake is manifest continuously throughout the wake evolution. This observation strengthens the results shown in previous sections (see Fig. 6-Fig. 8). While the momentum thickness for the non-stratified case is roughly constant throughout the wake, the momentum thickness of the stratified wake experiences major changes; it fluctuates from 0.5 to 2 chord lengths while slowly decreasing to a distance of 4 chord lengths, from which it increases with $x/c$ up to a distance of 6 chord lengths.

One can observe that the non-stratified normalized momentum thickness distribution for the early wake (see Fig. 9a) seems to overlap with the one of the late wake, (see Fig. 9b). Both triggering times (early and late) describe two phases in the wake evolution and it is not inconceivable to see them overlapped. The non-stratified wake depends on one time scale, where in the case of a constant accelerating hydrofoil the time can be directly related to the $x/c$ space (according to Taylor hypothesis). Therefore, one can expect the non-stratified wake to be self-similar; i.e., the non-stratified normalized momentum thickness distribution in the early wake should overlap with the one in the late wake. Fig. 9 clearly confirms the self-similarity of the average velocity profiles in the non-stratified wake.

Similar overlapping is not observed when the hydrofoil accelerated in the stratified layer, which enforces another time scale for the flow, namely the Brunt-Väisälä frequency, $N$. The presence of these two time scales, one related to the acceleration and another one to the stratification, explains that the stratification effect on the early (images captured after 3.2 sec) and late wakes (images captured after 5.1 sec) is quite different; i.e., the stratification effect on the hydrofoil wake is strongly related to time.

The lower momentum thickness values in the stratified case imply that the stratified layer deforms the vortices (as shown in Fig. 6), causing lower deficit in the wake behind the hydrofoil. We can deduce that the reduction of momentum thickness in the stratified case causes the wake height to be lower in comparison to the non-stratified wake. Such result is supported by the lower circulation values depicted in Fig. 8, but somewhat hard to determine from the vorticity maps (see Fig. 5-Fig. 6). One may assume that since conservation (mass, momentum and energy) is maintained, such reduction in the wake height must have been accompanied by some three-dimensional motion of the wake. In the following sub-section we make an assessment of that spanwise extent through an evaluation of the spanwise velocity gradient.

*3.4.1. 3D wake effects*

The growth of three-dimensional wakes in a stratified medium has been extensively studied in the literature (e.g., Spedding et al., 1996b; Meunier, 2012; Diamessis et al., 2011). In particular, the expansion of the spanwise width of a stratified wake was investigated. Spedding et al. (1996b) reported that the wake width of a towed sphere becomes narrower as Froude number decreases; i.e., as stratification effects become stronger. Thus, as buoyancy forces become more significant the three-dimensional turbulent wake behind the sphere becomes more two-dimensional. Yet, there are still some open questions regarding the stratification effect on a bluff body that originally generates a two-dimensional wake.

In the current study, the experimental setup is such that the flow about the hydrofoil symmetry plane can be treated as two-dimensional, due to the proximity of the side walls of the tank to the hydrofoil (see Fig. 1). This sort of experimental setup, where side walls bound the tested wing, was used extensively in the 1930's by NACA to characterize the aerodynamic coefficients of various two-dimensional airfoil sections (e.g., Jacobs et al., 1933). Yet, introducing the hydrofoil into a stratified layer may cause the wake to evolve in a different manner. To assess the three-dimensional effects in the wake behind the hydrofoil we calculate the spanwise velocity gradient, $\partial w/\partial z$, throughout the wake evolution pattern, which can provide some insight about the nature of the three-dimensional motion in the current experiments.

Starting from the continuity equation:

$$\nabla \cdot (\rho \vec{u}) = 0 \qquad (8)$$

The non-stratified wake has been evolved through a homogenous medium (constant density) and so Eq. (8) reduces to:

$$\frac{\partial w}{\partial z} = -\left(\frac{\partial u}{\partial x} + \frac{\partial v}{\partial y}\right) \qquad (9)$$

from which the spanwise velocity gradient can be easily deduced. For the stratified case, Eq. (8) can be reduced to the following form:

$$\frac{\partial w}{\partial z} = -\left(\frac{\partial u}{\partial x} + \frac{\partial v}{\partial y} + \frac{v}{\rho}\frac{d\rho}{dy}\right) \qquad (10)$$

where $v$ is the vertical velocity. Assuming $\rho \approx \rho_0$, we can rewrite the right term in Eq. (10) using the Brunt-Väisälä frequency, as follows:

$$\frac{\partial w}{\partial z} = -\left(\frac{\partial u}{\partial x} + \frac{\partial v}{\partial y}\right) + \frac{v}{g} N^2 \qquad (11)$$

One may notice the additional term in Eq. (11), $vN^2/g$, which is absent from the non-stratified case (see Eq. (9)).

The normalized spanwise velocity gradient, $(\partial w/\partial z) \cdot c/U_\infty$, was calculated for the non-stratified and stratified ensemble late wakes using Eq. (8) and Eq. (11), respectively. It should be noted that $\partial w/\partial z$ values were negligible in the non-stratified late wake. This result confirms that one may treat the wake pattern in the current study as two-dimensional when accelerating the hydrofoil through a non-stratified medium. Yet, this was not the case for the stratified late wake. High $\partial w/\partial z$ values were identified throughout the wake evolution, as shown in Fig. 10 that depicts the resultant normalized ensemble average spanwise velocity gradient for the stratified late wake. An additional normalized horizontal time axis, $at/U_\infty$, was added to Fig. 10, similar to the swirl function images shown in Fig. 7.

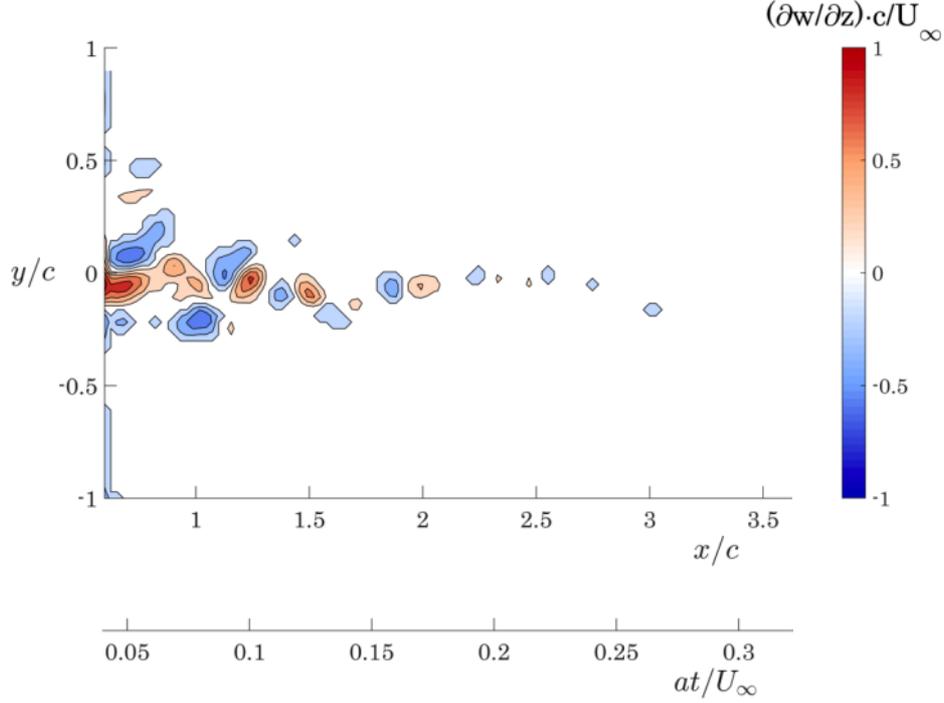

**Fig. 10.** Normalized ensemble average spanwise velocity gradient, $(\partial w/\partial z)\cdot c/U_\infty$, of the late wake behind a hydrofoil accelerating through a stratified layer ($Re_\infty$=5,000, $Fr_\infty$=2).

Fig. 10 depicts intense spanwise velocity gradients, characterized by a street of altering positive- and negative-signed values, which are generated throughout the wake evolution of the hydrofoil ($0.5 \leq x/c \leq 3$). The presence of this flow patterns in the spanwise direction are a result of an external forcing that the stratification impose on the flow. However, careful examination of the terms in eq. (11) shows that the buoyancy term, $vN^2/g$, is negligible in comparison to the velocity gradients. This result demonstrates that the stratification manifest the flow field resulting in three-dimensional effects. The strong $\partial w/\partial z$ gradients, which arise merely due to the stratification, could lead to some vortical motion that would transfer mass, momentum and energy in the spanwise direction, and generating a three-dimensional wake pattern in the hydrofoil symmetry plane that is absent in the non-stratified wake. These buoyancy originated gradients are presumably responsible for the high fluctuations in the momentum thickness of the stratified late wake (see Fig. 9b, at $0.5<x/c<1.5$).

The aforesaid results highlight the unique effect embedded in a stratified layer. The vertical density gradient introduces an additional buoyancy force that causes three-dimensional mixing of the wake behind any bluff body. This three-dimensional mixing is apparently absent in the wake generated in the non-stratified environment. As depicted in Fig. 10, although the non-stratified wake behind the symmetrical hydrofoil in the current study was shown to be two-dimensional, spanwise mixing can be observed in the stratified wake up to three chord lengths behind the hydrofoil.

### 3.5. *Phases in the wake of an accelerating bluff body in a stratified medium*

Following the aforesaid results, it is clear that a stratified layer strongly affects the wake pattern of an accelerating bluff body at low $Re$ and $Fr$. Characterizing the wake pattern through phase segmentation is essential in order to have a better understanding of the wake evolution behind an accelerating bluff body in a stratified medium; particularly when the stratified wake is characterized by two time scales (flow and stratification scales).

We begin our characterization by defining a non-dimensional time parameter, *Nt*, which will be used to describe the different wake phases. Again, *t* is defined as the 'wake age', as described in section 3.2.

As already shown in section 3.4, the momentum thickness is a good quantitative approach for assessing the stratification effect on the hydrofoil wake. Thus, the segmentation of the stratified wake into different phases will be accomplished using the normalized momentum thickness distribution throughout the wake with respect to the time, *t*.

Fig. 11 depicts the normalized momentum thickness distribution in the stratified medium, for the early and late wakes, with respect to the non-dimensional time, *Nt*. An additional normalized horizontal space axis, *x/c*, was also added to Fig. 11.

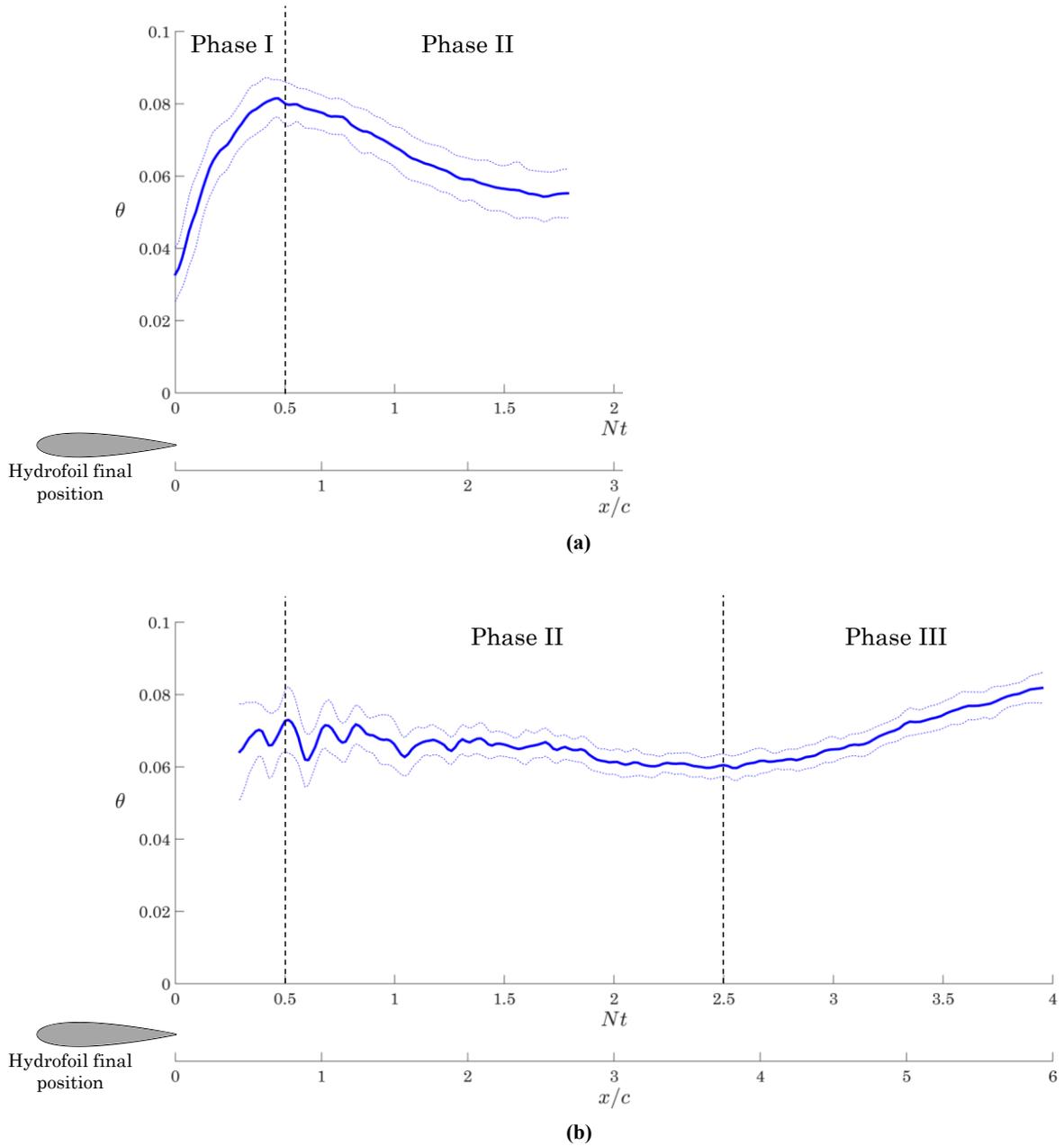

**Fig. 11.** Distribution of the normalized momentum thickness with *Nt* in the early ($Re_{80}$=4,000; $Fr_{80}$=1.6) and late ($Re_\infty$=5,000; $Fr_\infty$=2) wakes behind a hydrofoil accelerating through a stratified layer. Phases I, II and III are related to $0 \leq Nt \leq 0.5$, $0.5 \leq Nt \leq 2.5$ and $Nt \geq 2.5$, respectively.

As depicted in Fig. 11a, during phase I of the wake evolution, $0 \leq Nt \leq 0.5$, the momentum thickness of the stratified wake increases from the trailing-edge, and so the wake evolves and expands in the vertical direction similar to the non-stratified case (see Fig. 9); i.e., the wake expands in the *xy* plane behind the hydrofoil as a free shear flow (Kuethe and Chow, 1998). Phase I ends when the momentum thickness of the wake starts to decrease at $Nt \approx 0.5$, marking the start of the second phase (II), where the stratified layer starts affecting the hydrofoil wake, as shown by the velocity vectors in the vorticity maps of the early wake (see Fig. 5).

The momentum thickness distribution in the late stratified wake (see Fig. 11b) shows portions of the second and third phases (II-III). Phase I is not apparent since the PIV camera did not capture the entire near wake region; we assume this phase ended where the momentum thickness stops to increase at $Nt \approx 0.5$. During the second phase, the buoyancy forces first start to affect the wake, as depicted from the momentum thickness (see Fig. 10) and the absolute circulation (see Fig. 8a) distributions in the stratified late wake. At first, the momentum thickness of the wake fluctuates between $Nt \approx 0.5$ to $Nt \approx 1.0$. These fluctuations are remnants of the intense three-dimensional wake evolution shown by the spanwise velocity gradient distribution (see Fig. 10, at $0.5 < x/c < 1.5$). We may note that the spanwise mixing of the wake pattern is due to the disruption of the vertical density gradient in the tank by the hydrofoil, which immediately cause to three-dimensional buoyancy forces, accompanied with vortices being shed and suppressed (see Fig. 6-Fig. 7) throughout the wake. Yet, as the wake becomes older in age (*Nt* increases) its momentum dissipates and results to less spanwise mixing (see Fig. 10). We assume this is due to the stratified layer to resettle the disturbed vertical density gradient in the tank. As a result, buoyancy forces 'layered' the wake, thus causing a thinner momentum thickness, as depicted in Fig. 11b. It appears that this phenomenon is somewhat similar to the 'pancakes' vortices, which have been reported in the literature for higher *Nt* values (see Lin and Pao, 1979).

The slow decrease of the stratified late wake momentum thickness is depicted until $Nt \approx 2.5$. From $Nt \approx 2.5$ and further downstream the wake, the third phase (III) is visible; in that phase the effect of the stratified layer on flow pattern is apparently decreasing. The wake starts to expand again (similar to the initial phase) in the *xy* plane behind the hydrofoil, thus causing the momentum thickness to increase up to the same value depicted in the non-stratified wake (see Fig. 9).

Consequently, we can characterize the stratified wake behind the hydrofoil with two transitions; the first transition occurs at $Nt_I \approx 0.5$ from phase I to II and second transition occurs at $Nt_{II} \approx 2.5$ from phase II to III. These three phases, which depicted in Fig. 11 (I-III), characterize the stratified wake behind the accelerating hydrofoil in the current study. The initial effect of the stratified layer on the wake behind the accelerated hydrofoil is visible after half of the stratification time period, *T*; i.e., it takes half a cycle for the stratified layer to start and reshape the wake pattern. It appears that the strong effect of the stratified layer on the hydrofoil wake last for 2.5 cycles, after which buoyancy forces become weak and presumably not affect the stratified wake that resembles the non-stratified wake.

## 4. Conclusions

The present study characterizes the dynamics of a stratified wake, generated by a hydrofoil (NACA 0015) accelerating from rest at a constant rate. Throughout its acceleration the hydrofoil was subjected to low Reynolds and Froude numbers. PIV was utilized to extract the velocity fields in the wake region of the accelerated hydrofoil. The wake evolution of the hydrofoil has been characterized and a comparison is presented between the stratified and non-stratified (fresh water) wakes, in terms of the vorticity, circulation, spanwise velocity gradients and momentum thickness. Furthermore, we propose that the stratified wake of the hydrofoil during its acceleration stage can be divided into three phases that describe the effect of the external buoyancy force on the wake evolution.

The buoyancy force has shown a significant effect on the wake structure behind the accelerating hydrofoil. The stratification effect was observed continuously throughout the wake and was characterized by vortices that are being shed and stretched in the horizontal direction, thus resulting with lower spanwise vorticity values compared to those in the non-stratified wake. As a result, the stratified wake is characterized by lower circulation values compared to those resulted from the non-stratified medium. Moreover, the stratified late wake has shown smaller momentum thickness values in comparison to the non-stratified wake, which implies that the stratified wake is characterized with smaller velocity deficit. This is due to the three-dimensional effects in the stratified wake caused by the additional buoyancy force, which is absent in the non-stratified wake.

Three phases (I-III) were identified in the stratified wake yielding two transitions; the first transition occurs at $Nt_I \approx 0.5$ from phase I to II, where the stratification alters the wake pattern, and the second transition occurs at $Nt_{II} \approx 2.5$ from phase II to III, where the stratified wake resembles with the non-stratified wake. These phases characterize the stratified wake of the accelerating hydrofoil in the current study and show that stratification may have a strong effect on the wake, which could last for a short time period of 2.5 cycles. During the relatively short time period of the accelerated hydrofoil we have shown a reduction in the momentum thickness and some three-dimensional effects. These two phenomena allow the wake to dissipate into the spanwise direction, thus, causing a reduction of the drag necessary for the hydrofoil to overcome during its acceleration in the stratified medium.

Finally, this work might provide some guidance for underwater vehicles travelling in the pycnocline layer, for instance during commercial, research or military subsea exploration surveys. They are mounted with hydrofoils and tend to accelerate and decelerate, thus causing the wake pattern behind their hydrofoils to alter and influence their performance. Using high aspect ratio hydrofoils, underwater vehicles can gain a thinner momentum thickness and thus lower drag, as concluded in the current study. A thinner momentum thickness may have a high impact on surveillance missions, which required the underwater vehicle to be stealthiest as possible. Furthermore, a low drag vehicle should have a great significance in terms of energy consumption.

**Acknowledgments**

The authors wish to acknowledge the support from the Bi-National Science Foundation agency, grant #2008051.